\documentstyle[11pt,aaspp4]{article}

\def\xxx{\enspace\enspace\enspace}

\def\et{et al.}
\def\kms{km s$^{-1}$}
\def\ha{H$\alpha$}
\def\solar{\ifmmode_{\mathord\odot}\;\else$_{\mathord\odot}\;$\fi}

\def\HII{H$\,${\sc ii}}

\def\rh{R$_H$}

\def\hirat{R$_{HI}$/R$_{25}$}
\def\coldens{cm$^{-2}$}

\begin{document}

\title{The Distribution of Atomic Hydrogen Around Two Irregular Galaxies 
}

\author{Deidre A. Hunter}
\affil{Lowell Observatory, 1400 West Mars Hill Road, Flagstaff, Arizona 86001
USA;
\\dah@lowell.edu}

\and

\author{Eric M.\ Wilcots}
\affil{Washburn Observatory, University of Wisconsin, 475 N.\ Charter St.,
Madison, Wisconsin 53706 USA; \\ewilcots@uwast.astro.wisc.edu}

\begin{abstract}

We present radio interferometric observations of two irregular galaxies
that were candidates for having unusually extended HI emission. The galaxies,
UGC 199 and DDO 26, otherwise appeared to be normal, low-luminosity
systems with modest star-formation rates. 
To a detection limit of 2--3$\times10^{19}$ \coldens\ at a resolution
of about 50\arcsec, however,
the HI around neither galaxy is unusually extended compared to
other irregulars.
The HI around UGC 199 appears as a regular, symmetric distribution
with regular rotation and a maximum rotation speed of about 80 \kms.
By contrast, the HI around DDO 26 shows a concentration into two
blobs with an arm in the outer parts 
to the northwest and some additional emission
to the northwest of that. The kinematical 
major axis is approximately 75\arcdeg\ from the HI and optical
morphological axis which is unusual for Im galaxies.
In addition the velocity field in the outer parts of the galaxy
is messy and the velocity profiles at the two HI peaks are broad.
We suggest that DDO 26 has been perturbed externally or may be
two dwarfs in the process of merging.

\end{abstract}


\section{Introduction}

The atomic hydrogen gas in and around galaxies is the material out of
which star-forming clouds are derived and, hence, the HI plays a
crucial role in the evolution of galaxies. We have long been
intrigued by the gas that extends beyond, and sometimes well beyond,
the optical galaxy. In most irregulars, as in most spirals, the HI extends
to about twice the Holmberg radius \rh, defined as the radius at a
photographic surface brightness level of 26.7 magnitudes per arcsec$^2$ 
(see Figure 13 of Hunter [1997] and references therein for a compilation
of HI extents). However, some galaxies have gas extending as far as
7\rh. 

In order to determine the role of this extended gas in the life of
irregular galaxies, we began a radio interferometric study of 
the gas around those
with particularly large HI extents at high enough resolution to
see interesting structure if it were present.
The general questions we are addressing are: 
1) What is the structure of extended gas disks
around irregular galaxies; 
2) What role does this potential gas reservoir play in
the evolution of
the galaxies; and
3) Can large primordial gas disks around irregulars survive without
disruption?  These questions are important to understanding the internal
processes and evolution of irregulars, which are
the most numerous type of star-forming galaxy.   
Additionally, if, as studies of damped
Ly$\alpha$ systems indicate, the majority of galaxies were 
larger HI clouds in their youth
(Rao, Turnshek, \& Briggs 1995), nearby irregulars with large gas envelopes 
are local examples of high redshift
damped Ly$\alpha$ systems, and this has been substantiated
with observations of one such irregular that lies directly
in front of a QSO (Bowen, Tripp, \& Jenkins 2001).
Understanding the state of 
these envelopes helps us understand the
process of galaxy formation.

Two of the galaxies whose extended gas disks we have studied
so far turned out to have HI envelopes that are
most definitely {\it not} regular, quiescent disks.
NGC 4449 is nearly
surrounded by a large arc of gas that spans roughly 80 kpc
(Hunter \et\ 1998).  Interestingly, not only is
this gas cold (velocity dispersions of 5--10 km s$^{-1}$), 
but it is also in regular
rotation about the center of the galaxy.
HI maps of IC 10 show that the extended gas is concentrated
in three arm-like structures, and that
IC 10 is merging with a large infalling cloud currently 
to the south of the disk
(Wilcots \& Miller 1998). A third galaxy, Sextans A, by contrast,
does appear to have a smooth outer HI envelope (Wilcots \& Hunter 2001).

To increase our sample, we obtained observations of two galaxies
that were candidates for having unusually extended HI gas:
UGC 199 and
DDO 26 ($=$UGC 2053).
UGC 199 was chosen from the Arecibo Observatory mapping of
van Zee, Haynes, \& Giovanelli (1995).
DDO 26 was chosen from the list of Hunter \& Gallagher (1985)
that was derived from comparing HI fluxes obtained with
single-dish radio telescopes with different beam sizes: 
the NRAO 140-foot and the NRAO 300-foot telescopes with beam
sizes of 21\arcmin\ and 10\arcmin, respectively.
We present the results of new interferometric
observations of these two galaxies
here, and show that, in fact, neither of them is likely to have unusually
extended HI emission.

Properties of UGC 199 and DDO 26 are given in Table \ref{tabgal}.
Both systems are classed as Im galaxies and were not known 
to be interacting systems. DDO 26 has an M$_B$ that
is comparable to that of the SMC. We do not yet have an M$_B$ for
UGC 199, but an estimate by van Zee \et\ (1995) from a photographic
magnitude would place UGC 199 as 2 magnitudes fainter than DDO 26.
The star formation rates derived from \ha\ luminosities 
are also given in Table \ref{tabgal}. DDO 26 has a star formation
rate per unit area that is between that of the SMC and LMC.
UGC 199's star formation rate is 16 times lower and is typical
of irregulars (Hunter 1997). 
Thus, both objects are relatively low luminosity and appear to
be fairly typical Im
galaxies.

\section{Observations}

\subsection{HI}

We obtained 21 cm line-emission observations of UGC 199 and DDO 26 with the
Very Large Array
(VLA\footnote{
The VLA is a facility of the National Radio Astronomy Observatory (NRAO),
itself a facility of the National Science Foundation that is operated
by Associated Universities, Inc.})
radio interferometer
in its D-array configuration on 5 August 2000.
We chose the D-array configuration in order to maximize sensitivity
to low column density extended emission.
Characteristics of the observations and maps are given in Table \ref{tabhi}.
The total bandwidth was
1.56 MHz with 128 channels and a channel separation of 12.2 kHz (2.6 \kms).
The data were on-line Hanning-smoothed, and the resulting velocity
resolution is 5 \kms.

We subtracted the continuum emission in the {\it uv}-plane using
line-free channels on either end of the spectrum.
To make maps, we employed a routine in NRAO's
Astronomical Image Processing System (AIPS)
that enables one to choose a sample weighting that is
in between the standard ``natural'' weighting, which gives the
highest signal-to-noise but at the expense of long wings in the
beam, and ``uniform'' weighting, which gives the best resolution
but at the cost of signal-to-noise and the presence of negative
sidelobes.
We chose a sample weighting
that gives a formal increase
of only 5\% in noise yet with a significant improvement in beam
profile over what would have been achieved with
``natural'' weighting.  The resulting synthesized beam profiles
(FWHM) are given in Table \ref{tabhi}.
We also experimented with natural-weighted maps and with
maps smoothed to twice the beam-size in order to explore 
possible missed emission at low column densities.

We deconvolved the
maps until there were roughly comparable numbers of positive and
negative components.
To remove the portions of each channel map without emission that
are only contributing noise to the map, we used
the maps smoothed to twice the beamsize for a conditional transfer of
data in the unsmoothed maps. The channel maps were blanked wherever
the flux in the smoothed maps fell below 2.5$\sigma$.
Flux-weighted moment maps were made from the resulting data cube.

\subsection{V-band Images}

We obtained V-band images of UGC 199 with the Perkins 1.8 m telescope
at Lowell Observatory 1998 December and 1999 January. We used
a SITe 2048$\times$2048 CCD binned 4$\times$4. The pixel scale
was 0.61\arcsec\ and the seeing was 2.2\arcsec. The night was clear.
We took three 750 s exposures and combined them to remove cosmic rays.

DDO 26 was observed by P.\ Massey
in V-band with the Kitt Peak National Observatory
4 m telescope 1997 October. The exposure was a single 60 s image.
The detector was a Tektronix 2048$\times$2048 CCD. The night was clear.
The pixel scale was 0.42\arcsec\ and the seeing was 1.6\arcsec.

The electronic pedestal was subtracted using the overscan strip.
Images were flat-fielded using observations of the twilight sky.

\section{The HI Results}

\subsection{UGC 199}

To examine continuum objects in the field of view of the radio map
of UGC 199,
we constructed and deconvolved maps from the D-array
{\it uv}-data before the
continuum was subtracted.
The 21 cm continuum emission
is illustrated in Figure \ref{figcont1}. An outer HI contour
of UGC 199 is shown superposed to outline the galaxy. We have
not detected any continuum emission from UGC 199 itself.
Other sources in the field of view of the primary beam are listed 
in Table \ref{tabu199cont} along with their fluxes.

Channel maps of the HI line-emission in UGC 199 are shown in
Figure \ref{figchan1}. We detected HI from 
1740 \kms\ to 1870 \kms. The channel maps show an elliptical
distribution that does not change significantly from channel to
channel and that appears to be similar to the beam shape.
There is clear evidence of regular rotation.

The integrated HI map is shown superposed on our V-band image
in Figure \ref{figm01}. The distribution is smooth and symmetrical.
It is elongated 
at a position angle of $-$77\arcdeg, close to that of the optical.
The minor-to-major axis ratio of the HI is 0.77. If the intrinsic
ratio is 0.3, as found in the optical for other irregulars
by Hodge \& Hitchcock (1966) and van den Bergh (1988),
the observed ratio implies an inclination of 42\arcdeg.

The flux in each channel of the HI
data cube was integrated over
a square 21\arcmin\ on a side
in order to determine the total
HI flux detected in the map.
The flux was corrected
for attenuation by the primary beam.
The integrated profile is shown in Figure \ref{figsingle1}
and compared to the single-dish profile
obtained by Schneider \et\ (1990). The beam of that observation
was 3.3\arcmin, and so it is no surprise that our VLA observation
detects more gas. 
We detect a total HI mass of 
$8.5\times10^8$ M\solar. For comparison 
van Zee \et\ (1995) give a total HI mass of $8.1\times10^8$ M\solar.
Thus, our VLA map includes 5\% more HI.
Maps of our data made with lower resolution but higher signal-to-noise
do not detect any more flux.

Velocity field contours are shown superposed on our V-band image
in Figure \ref{figm11}. There is clear and regular rotation
at a position angle of $-$82\arcdeg, close to the major axis of the 
integrated HI distribution. 
We have determined a rotation curve in the following manner.
We began by allowing all parameters to be variables
and fitting the entire field with a Brandt function. From the
resulting solution we fixed the center coordinates at
00$^h$ 20$^m$ 50.9$^s$, 12\arcdeg\ 51\arcmin\ 39\arcsec\ (epoch 2000).
We then fit the inner 40\arcsec\ radius of the velocity field
with solid body rotation and from that fixed the systemic velocity
at 1800.5$\pm2.0$ \kms. This agreed to 0.3 \kms\ with the
systemic velocity determined from the fit with the Brandt function as
well.
We then fit tilted ring models to annuli of 20\arcsec\ width.
Here the position angle was fixed at $-$82\arcdeg\ from Figure \ref{figm11}
and the inclination at 42\arcdeg, as discussed above.
The resulting rotation curve is shown in Figure \ref{figrot1}.
We see UGC 199 attains a rotation speed of about 80 \kms,
for an inclination of 42\arcdeg.

\subsection{DDO 26}

To examine continuum objects in the field of view of the radio map
of DDO 26,
we constructed and deconvolved maps from the D-array
{\it uv}-data before the
continuum was subtracted.
The 21 cm continuum emission
is illustrated in Figure \ref{figcont2}. An outer HI contour
of DDO 26 is shown superposed to outline the galaxy. We have
not detected any continuum emission from DDO 26 itself.
Other sources in the field of view of the primary beam are listed 
in Table \ref{tabd26cont} along with their fluxes.

Channel maps of the HI line-emission in DDO 26 are shown in
Figure \ref{figchan2}. We detected HI from 
974 \kms\ to 1078 \kms.
There is clear division into two emission peaks from 977 \kms\ to
1039 \kms.
The channel maps also show an arm to the northwest of the galaxy center
from 1031 \kms\ to 1068 \kms. We also display the integrated
HI map in Figure \ref{figarm} so as to bring out the arm.

The integrated HI map is shown superposed on our V-band image
in Figure \ref{figm02} and on the \ha\ image in Figure \ref{figm22}.
The inner HI distribution is elongated along
a position angle of 46\arcdeg. This is rotated about 8\arcdeg\
further to the east than the apparent position angle of the optical
distribution that has a major axis position angle of 38\arcdeg\
in the V-band (see also Swaters 1999).
The outer HI contour is close to round even though the center of
the HI distribution and the optical are clearly elongated.
The second lowest contour, on the other hand, is more boxy than round.
These changes in minor-to-major axis ratio with radius could indicate
changes in inclination of the gas disk and be the result of
a warp.
Warps in gas disks are a common phenomenum, but usually in normal
systems they are not as extreme as 
what we see here.

The integrated HI shows the two peaks with centers about 41\arcsec\
apart. The \HII\ regions are clustered around these two peaks.
The arm seen in the channel maps is apparent as a
bulging of the contours to the northwest of the galaxy center.
There is also a piece of emission that appears to the 
northwest of the arm in channels 1057 \kms\ to 1075 \kms\
and appears in the integrated HI map as a knob on the outer 
contour.

The flux in each channel of the HI
data cube was integrated over
a square 21\arcmin\ on a side
in order to determine the total
HI flux detected in the map.
The flux was corrected
for attenuation by the primary beam.
The integrated profile is shown in Figure \ref{figsingle2}
and compared 
to the NRAO 140-foot Telescope single-dish
observation of Hunter \& Gallagher (1985) that used a beam of 21\arcmin.
In Figure \ref{figsingle2} one can see that the 140-foot telescope and
VLA profiles parallel each other, but the 140-foot profile is 
everywhere significantly higher. 
We detect a total HI mass of 
$1.1\times10^9$ M\solar\ while Hunter and Gallagher
detected $2.0\times10^9$ M\solar. Thus,
we would appear to have detected only 56\% of the HI
detected with a single-dish telescope. We will return to this issue
below.

Velocity field contours are shown superposed on our V-band image
in Figure \ref{figm12}. The velocity field is fairly regular
in the center, becoming disorganized in the outer parts.
However, the major axis implied by the velocity field is
about $-$29\arcdeg\ (Swaters 1999 measured $-$37\arcdeg). 
This is 75\arcdeg\ to the northwest of the position angle
defined by the major axis of the inner HI distribution.
Thus, the rotation axis is quite divorced from the morphology
of the HI and optical.

How common is such a large difference in the position angle
of the kinematic and morphological major axes in Im galaxies?
To address this we examined the sample of Im and Sm galaxies
observed by Swaters (1999). We selected from his sample those
galaxies with organized velocity fields and with ellipticities
greater than 0.1 so that the optical major axis would not be
ambiguous. These criteria left 47 galaxies from his larger sample.
We determined the difference between the position angles of
his HI kinematical axis and his R-band optical morphological
axis. We checked most galaxies and adjusted three position
angles: one obvious typographical error (UGC 6446) and
two that disagreed by large amounts with RC3 values and with
our own V-band image (UGC 10310, UGC 11861). The resulting
number distribution is shown in Figure \ref{figpa}. The position
of DDO 26 is marked. One can clearly see that most Im and Sm
galaxies have morphologies and velocity fields that agree
within 10\arcdeg, although there is a tail in the distribution
up to 40\arcdeg\ difference. Only three galaxies, including DDO 26,
have differences larger than this, and all three have differences
of 60\arcdeg---70\arcdeg. Thus, we see that while DDO 26 is not alone in
exhibiting such a large difference, it is unusual compared to most
Im and Sm galaxies.

We show a position-velocity diagram in Figure \ref{figlp2}, for
slices one beam-width wide 
along the optical major axis and the HI kinematic axis.
The slice along the optical major axis reflects the two HI peaks
and the intensity minimum between them. Otherwise the two slices
show a very broad range in velocity at a given location.
Figure \ref{figblobs2} shows profile cuts at the locations of the
two HI peaks, integrated over a square approximating the beam-size.
One can again see the wide velocity width (44 \kms\ and
51 \kms\ FWHM). The higher intensity
peak (labeled 1 in the figure) has a shallow fall-off to higher
velocities and the lower intensity peak (labeled 2 in the figure)
appears to be the blending of two Gaussians.

Clearly, DDO 26 is not kinematically relaxed.  The
dramatic change in the position angle of the HI distribution, the
large offset between the kinematical and morphological major axes,
the messy kinematics in the outer parts of the velocity field,
the broad profiles of the HI peaks, and the HI arm to the
northwest combine to suggest that 
this galaxy has been perturbed sometime in the recent past.
There is no obvious perturbing galaxy nearby; the nearest large
galaxy (NGC 1012) is 330 kpc on the plane of the sky and 42 \kms\
different in radial velocity.
However, the double peaked nature of the HI and optical suggest
another possibility: that DDO 26 is two systems in the process
of merging.
In this scenario, the arm to the northwest would be the
result of tidal forces.
It is not always easy to tell galaxies that are irregular
due to internal processes from 
galaxies that are irregular due to external processes.
And certainly, normal Im galaxies are lumpy in the HI and optical.
However, given the unusual characteristics of this system,
a merging system is a possibility.
If DDO 26 is two systems
in the process of becoming one, the original systems must both
have been small irregular galaxies to begin with
because the mass and luminosity of the combined systems are
low.

\section{HI Extents}

\subsection{UGC 199}

For UGC 199 we measure a maximum HI radius of 1.6\arcmin\
at the outer contour of 2.2$\times10^{19}$ \coldens\ in 
Figure \ref{figm01}. Thus, the HI extent relative to the optical
\hirat\ is only 3.6. 
Estimating R$_H$ as 40\% larger than R$_{25}$, we have R$_{HI}$/R$_H\sim2.5$.
From data compiled from the literature for Im galaxies, Hunter (1997)
shows a peak in the distribution of R$_{HI}$/R$_H$ of 1.5---2. 
Note, however, that Hunter's R$_{HI}$ is the extent measured to
$1\times10^{19}$ \coldens.
The HI around DDO 26, therefore, is only marginally more extended than
that of typical irregulars. 
There would have to be highly extended HI around UGC 199 at
column densities 1---2$\times10^{19}$ \coldens\ for the extent
of the HI around UGC 199 to be much larger. 
Since our smoothed data which reaches this sensitivity limit does
not pick up any more flux, this possibility seems unlikely.

\subsection{DDO 26}

For DDO 26 we measure a maximum HI radius of 2.4\arcmin\
at the outer contour of 3.1$\times10^{19}$ \coldens\
in Figure \ref{figm02}. Thus, \hirat\ is 2.4.
From Swaters' (1999) R-band d$_{25}$/d$_{26.5}$ ratio for DDO 26,
we estimate R$_{HI}$/R$_H\sim1.7$, a value that is typical for Im
galaxies. Thus, there would have to be considerable gas
1--3$\times10^{19}$ \coldens\ at large radii for DDO 26 to have unusually
extended HI.

On the other hand, the comparison with Hunter \& Gallagher's
(1985) single-dish observations says that we have only detected 56\% of
the gas emission. Furthermore, Hunter and Gallagher would predict
an HI diameter greater than 10\arcmin, a factor of two larger
than what we detect.
There are three possible explanations.
First, the extended emission is there but it 
is larger than the largest structure
the VLA D-array is sensitive to and is absolutely smooth beyond
our detection radius. This would require that the missing HI be
extended $>$15\arcmin\ radius. 
Since the two beam-sizes in Hunter and Gallagher's 
flux comparison were 10\arcmin\ and 21\arcmin, this is possible.

Second, the missing emission is at a column density
below our detection limit of $<3\times10^{19}$.
In the worst case scenario, the missing emission---9$\times10^8$ 
M\solar---would be spread evenly over an annulus of radius from
2.4\arcmin, the extent of what we detect, to 10.5\arcmin, the half beam
size of the Hunter \& Gallagher (1985) observation. That would
be a column density of 1$\times10^{19}$ \coldens. 
Although Figure \ref{figm02} only goes to 3$\times10^{19}$ \coldens,
we would easily have detected a column density of 1$\times10^{19}$ \coldens\
in our smoothed map, and we do not see this emission.

The third, explanation is that
Hunter \& Gallagher (1985) goofed and, in fact, the HI emission
associated with DDO 26 is not highly extended. That this might
be the case is suggested by Figure \ref{figsingle2} where the
Hunter and Gallagher integrated profile matches our VLA profile
closely in shape but the Hunter and Gallagher profile is offset
to higher flux values. If the Hunter and Gallagher profile is
sitting on a 0.03 Jy pedestal, a 9\% error in the peak,
the two profiles could be brought into agreement.
If we were simply missing extended HI that Hunter and Gallagher
had detected, we might not expect that missing flux to be so
evenly distributed in velocity as Figure \ref{figsingle2} suggests.
Thus, this explanation seems the most likely. 

\section{Summary}

We have presented VLA D-array observations of two irregular galaxies
that were candidates for unusually extended HI emission. The galaxies,
UGC 199 and DDO 26, otherwise appeared to be normal, low-luminosity
systems with modest star-formation rates.
Our VLA data suggest that the HI around these two galaxies is not
unusually extended.

The HI around UGC 199 appears as a regular, symmetric distribution
with regular rotation and a maximum rotation speed of about 80 \kms.
We detect HI to 1.6\arcmin\ at 2.2$\times10^{19}$ \coldens. The
ratio \hirat\ is 3.4 which is similar to 
values for other irregulars.

By contrast, the HI around DDO 26 shows a concentration into two
blobs with an arm to the northwest and some additional emission
to the northwest of that. The major axis of rotation is 75\arcdeg\
from the morphological major axis, and this is unusual for Im and Sm galaxies.
We suggest that DDO 26 has been perturbed in some way and that it
could be
two dwarfs in the process of merging.
We detect HI to a radius of 2.4\arcmin\ at 3$\times10^{19}$ \coldens.
\hirat\ is 2.4 which is typical of Im galaxies.
However, we find that we are missing 56\% of the flux detected in
a single-dish observation and conclude that most likely the 
single-dish
observation is in error.

\acknowledgments

We wish to thank the staff at the VLA who provided the support
that made these observations possible.
This work was supported
by the National Science Foundation under Grant No.\
AST-9616940 to DAH and AST-9616907 to EMW and the Lowell Research
Fund.

\clearpage
 
\begin{deluxetable}{lccc}
\tablecaption{Galaxy Characteristics. \label{tabgal}}
\tablewidth{0pt}
\tablehead{
\colhead{Quantity} & \colhead{UGC 199}   & \colhead{DDO 26}
& \colhead{Reference}
} 
\startdata
Morphological type & Im: & Im & RC3 \nl
D (Mpc) & 29.8 & 17.3 & RC3, H$_0=65$ km s$^{-1}$ Mpc$^{-1}$ \nl
R$_{25}$ (arcmin) & 0.46 & 1.02 & RC3 \nl
R$_{25}$ (kpc)    & 7.9 & 5.15 & RC3 \nl
E(B$-$V)$_f$ & 0.038 & 0.093 & Burstein \& Heiles 1984 \nl
E(B$-$V)$_t$\tablenotemark{a} & \nodata & 0.50 & Hunter \& Hoffman 1999 \nl
M$_B$ & \nodata & $-$16.68 & RC3 \nl
log L$_{H\alpha}$ (ergs s$^{-1}$) & 39.23 & 39.50 & Hunter \& Elmegreen 2001 \nl
log SFR/area (M\protect\solar yr$^{-1}$ kpc$^{-2}$)\tablenotemark{b} 
& $-$3.61 & $-$3.57 
& Hunter \& Elmegreen 2001 \nl
\enddata
\tablenotetext{a}{E(B$-$V)$_t$ is used to correct the \protect\ha\ 
luminosity for reddening, in conjuntion with the reddening law of
Cardelli, Clayton, \& Mathis (1989). For UGC 199, we assumed
an E(B$-$V)$_t$$=$E(B$-$V)$_f$+0.1.}
\tablenotetext{b}{The star formation rate per unit area
(SFR/area) is determined from the \protect\ha\ luminosity and
a Salpeter (1955) stellar initial mass function integrated from
0.1 M\protect\solar\ to 100 M\protect\solar\ (Hunter \& Gallagher 1986).
The area is $\pi$R$_{25}^2$.}
\end{deluxetable}

\clearpage
 
\begin{deluxetable}{lcc}
\tablecaption{HI Observations and Data. \label{tabhi}}
\tablewidth{0pt}
\tablehead{
\colhead{Quantity} & \colhead{UGC 199}   & \colhead{DDO 26}
} 
\startdata
RA (2000) & 00 20 51.4 & 02 34 29.4 \nl
DEC (2000) & 12 51 39 & 29 45 05 \nl
Central velocity (\protect\kms) & 1800 & 1034 \nl
Time on source (minutes) & 140 & 145 \nl
Beam FWHM (arcsec) & 55.2$\times$46.9 & 51.0$\times$45.5 \nl
Beam FWHM (kpc) & 8.0$\times$6.8 & 4.3$\times$3.8 \nl
Single channel rms (Jy beam$^{-1}$) & 0.94 & 0.96 \nl
Integrated HI (M\protect\solar) & 8.5$\times10^8$ & 1.1$\times10^9$ \nl
R$_{HI}$ to $3\times10^{19}$ cm$^{-2}$ (arcmin) & 1.5 & 2.4 \nl
R$_{HI}$/R$_{25}$  & 3.3 & 2.4 \nl
\enddata
\end{deluxetable}

\clearpage
 
\begin{deluxetable}{rccc}
\tablecaption{Continuum sources near UGC 199. \label{tabu199cont}}
\tablewidth{0pt}
\tablehead{
\colhead{Object} & \colhead{RA (2000)\tablenotemark{a}}   
& \colhead{DEC (2000)\tablenotemark{a}}
& \colhead{Flux (Jy)}
} 
\startdata
1\xxx & 00 20 55.8 & 13 13 32 & 0.232 \nl
2\xxx & 00 21 41.2 & 13 10 30 & 0.014 \nl
3\xxx & 00 20 10.5 & 13 10 43 & 0.108 \nl
4\xxx & 00 21 58.1 & 13 03 07 & 0.019 \nl
5\xxx & 00 19 51.8 & 13 02 03 & 0.014 \nl
6\xxx & 00 19 37.6 & 13 02 41 & 0.013 \nl
7\xxx & 00 19 44.7 & 12 58 21 & 0.038 \nl
8\xxx & 00 19 46.5 & 12 54 53 & 0.026 \nl
9\xxx & 00 22 31.0 & 12 57 03 & 0.030 \nl
10\xxx & 00 20 28.3 & 12 52 57 & 0.078 \nl
11\xxx & 00 20 36.3 & 12 47 32 & 0.017 \nl
12\xxx & 00 20 33.6 & 12 46 40 & 0.007 \nl
13\xxx & 00 20 33.6 & 12 43 25 & 0.031 \nl
14\xxx & 00 20 16.8 & 12 37 08 & 0.016 \nl
15\xxx & 00 19 47.5 & 12 28 15 & 0.092 \nl
16\xxx & 00 21 44.7 & 12 33 01 & 0.021 \nl
\enddata
\tablenotetext{a}{Units of RA are hours, minutes, and seconds.}
\tablenotetext{b}{Units of DEC are degrees, arcminutes, and arcseconds.}
\end{deluxetable}

\clearpage
 
\begin{deluxetable}{rccc}
\tablecaption{Continuum sources near DDO 26. \label{tabd26cont}}
\tablewidth{0pt}
\tablehead{
\colhead{Object} & \colhead{RA (2000)\tablenotemark{a}}   
& \colhead{DEC (2000)\tablenotemark{a}}
& \colhead{Flux (Jy)}
} 
\startdata
1\xxx & 02 34 36.4 &  29 57 52 & 0.005 \nl
2\xxx & 02 33 40.4 &  29 56 33 & 0.007 \nl
3\xxx & 02 35 00.4 &  29 50 17 & 0.004 \nl
4\xxx & 02 35 03.3 &  29 46 23 & 0.005 \nl
5\xxx & 02 33 31.5 &  29 47 27 & 0.005 \nl
6\xxx & 02 33 44.5 &  29 42 54 & 0.026 \nl
7\xxx & 02 34 19.4 &  29 33 36 & 0.020 \nl
8\xxx & 02 33 55.5 &  29 32 44 & 0.008 \nl
9\xxx & 02 33 37.6 &  29 30 46 & 0.004 \nl
10\xxx & 02 34 58.3 &  29 29 42 & 0.006 \nl
\enddata
\tablenotetext{a}{Units of RA are hours, minutes, and seconds.}
\tablenotetext{b}{Units of DEC are degrees, arcminutes, and arcseconds.}
\end{deluxetable}

\clearpage

\figcaption{21-cm continuum map of the region around UGC 199
is shown.
The map has been corrected for the attenuation by the primary beam.
The continuum sources are numbered as given in Table \protect\ref{tabu199cont}.
The contour near the center is the outer HI-emission contour of
UGC 199 shown in Figure \protect\ref{figm01}, and is
2.2$\times10^{19}$ cm$^{-2}$.
The beam size FWHM (55.2\protect\arcsec$\times$46.9\protect\arcsec)
is shown as the elliptical contour in the lower left corner.
\label{figcont1}}

\figcaption{Channel maps of HI line-emission in UGC 199 made from
VLA D-array observations. The inner 
5.6\protect\arcmin$\times$5.6\protect\arcmin\ are shown.
Every other channel from the data cube is shown. 
The beam size FWHM (55.2\protect\arcsec$\times$46.9\protect\arcsec)
is shown in the final panel.
Right Ascension is marked along the x-axis. The right tick marks
an RA of 00$^h$ 20$^m$ 45$^s$; the left tick marks
an RA of 00$^h$ 21$^m$ 00$^s$.
\label{figchan1}}

\figcaption{Integrated HI map of UGC 199 is shown as contours 
superposed on our V-band image of the galaxy.
The HI contour levels are 
2.2, 3.3, 14.5, 25.7, 36.9, 48.0, and 59.2$\times10^{19}$ cm$^{-2}$.
The beam size FWHM (55.2\protect\arcsec$\times$46.9\protect\arcsec)
is shown as the elliptical contour in the lower left corner.
\label{figm01}}

\figcaption{Integrated HI profile of UGC 199 from our VLA D-array
data. For comparison we also show the integrated profile from 
Schneider et al.'s (1990)
observations of UGC 199 taken with the Arecibo Observatory and
a 3.3\arcmin\ beam.
\label{figsingle1}}

\figcaption{Contours of the velocity field of UGC 199 are shown
superposed on our V-band image.
The contours are 1750 \protect\kms\ to 1840 \protect\kms\ in
steps of 10 \protect\kms. The first and last contours are labeled.
The beam size FWHM (55.2\protect\arcsec$\times$46.9\protect\arcsec)
is shown as the elliptical contour in the lower left corner.
\label{figm11}}

\figcaption{Best-fit rotation curve of UGC 199. The velocity field
was fit in annuli 20\protect\arcsec\ in width spaced every 
20\protect\arcsec. We held the position angle fixed at $-$82\protect\arcdeg.
This was determined from Figure \protect\ref{figm11}.
We also held the inclination angle fixed at 42\protect\arcdeg. This
was determined from the minor-to-major axis ratio of the integrated
HI (Figure \protect\ref{figm01}) and assumed an intrinsic ratio 
of 0.3 (Hodge \& Hitchcock 1966, van den Bergh 1988).
Determination of the central position and central velocity 
are discussed in the text.
\label{figrot1}}

\figcaption{21-cm continuum map of the region around DDO 26
is shown.
The map has been corrected for the attenuation by the primary beam.
The continuum sources are numbered as given in Table \protect\ref{tabd26cont}.
The contour near the center is the outer HI-emission contour of
DDO 26 shown in Figure \protect\ref{figm02}, and is
3.1$\times10^{19}$ cm$^{-2}$.
The beam size FWHM (51.0\protect\arcsec$\times$45.5\protect\arcsec)
is shown as the elliptical contour in the lower left corner.
\label{figcont2}}

\figcaption{Channel maps of HI line-emission in DDO 26 made from
VLA D-array observations. The inner 
9.75\protect\arcmin$\times$9.75\protect\arcmin\ are shown.
The beam size FWHM (51.0\protect\arcsec$\times$45.5\protect\arcsec)
is shown in the final panel.
Right Ascension is marked along the x-axis: The right tick marks
an RA of 02$^h$ 34$^m$ 15$^s$; the middle tick marks
an RA of 02$^h$ 34$^m$ 30$^s$; and the
left tick marks 
an RA of 02$^h$ 34$^m$ 45$^s$.
\label{figchan2}}

\figcaption{Integrated HI map of DDO 26 shown so as to bring out
the arm to the northwest of the center of the galaxy.
\label{figarm}}

\figcaption{Integrated HI map of DDO 26 is shown as contours 
superposed on our V-band image of the galaxy.
The HI contour levels are 
3.1 12.5, 25.0, 37.5, 50.0, 62.5, 75.0, 87.5, 100.0, 112.5,
125.0, and 137.5$\times10^{19}$ cm$^{-2}$.
The beam size FWHM (51.0\protect\arcsec$\times$45.5\protect\arcsec)
is shown as the elliptical contour in the lower left corner.
\label{figm02}}

\figcaption{Integrated HI map of DDO 26 is shown as contours 
superposed on our H$\alpha$ image of the galaxy.
The stellar continuum has been subtracted from the H$\alpha$ image
to leave only nebular emission.
The HI contour levels are 
3.1 12.5, 25.0, 37.5, 50.0, 62.5, 75.0, 87.5, 100.0, 112.5,
125.0, and 137.5$\times10^{19}$ cm$^{-2}$.
The beam size FWHM (51.0\protect\arcsec$\times$45.5\protect\arcsec)
is shown as the elliptical contour in the lower left corner.
\label{figm22}}

\figcaption{Integrated HI profile of DDO 26 from our VLA D-array
data. For comparison we also show the integrated profile from 
Hunter \& Gallagher's (1985) observations using the NRAO 140-foot
radio telescope which has a 21\arcmin\ beam.
\label{figsingle2}}

\figcaption{Contours of the velocity field of DDO 26 are shown
superposed on our V-band image.
The contours are 1010 \protect\kms\ to 1050 \protect\kms\ in
steps of 5 \protect\kms. The first and last two contours are labeled.
The beam size FWHM (51.0\protect\arcsec$\times$45.5\protect\arcsec)
is shown as the elliptical contour in the lower left corner.
\label{figm12}}

\figcaption{Differences between optical and HI kinematical position
angles for a sample of 47 Im and Sm galaxies taken from Swaters (1999).
We have marked the value for DDO 26. 
\label{figpa}}

\figcaption{Position-velocity plots for slices through the DDO 26
HI data cube. The top panel is a slice at the position angle of
the optical major axis; the bottom panel, along the HI kinematic
major axis. Each slice is a sum over one beam width through the
center of the HI map.
The contours in the top panel go from 0.025 to 0.2 in steps of 0.025 Jy
beam$^{-1}$; in the bottom panel, from 0.025 to 0.25 in steps of 0.025
Jy beam${-1}$.
\label{figlp2}}

\figcaption{Profile cuts through the two HI peaks seen in 
Figure \protect\ref{figm02}. For each we have integrated over a square
with a side equal approximately to the beam-size. 
Peak 1 is centered at 
2$\rm^h$ 34$\rm^m$ 28.4$\rm^s$, 29\arcdeg\ 44\arcmin\ 39\arcsec;
peak 2 is centered at
2$\rm^h$ 34$\rm^m$ 31.4$\rm^s$, 29\arcdeg\ 45\arcmin\ 18\arcsec.
\label{figblobs2}}


\clearpage

\begin{references}

\reference{} Bowen, D.\ V., Tripp, T.\ M., \& Jenkins, E.\ B.\ 2001, AJ,
in press
\reference{} Burstein, D., \& Heiles, C. 1984, ApJS, 54, 33
\reference{} Cardelli, J.\ A., Clayton, G.\ C., \& Mathis, J.\ S.\
1989, ApJ, 345, 245
\reference{} de Vaucouleurs, G., de Vaucouleurs, A., Corwin, H., Buta, R.,
Paturel, G., \& Fouqu\'e, P.\ 1991, Third Reference Catalogue of Bright
Galaxies (New York, Springer-Verlag) (RC3)
\reference{} Hodge, P.\ W., \& Hitchcock, J.\ L.\ 1966, PASP, 78, 79
\reference{} Hunter, D.\ A.\ 1997, PASP, 109, 937
\reference{} Hunter, D.\ A., \& Elmegreen, B.\ G.\ 2001, in preparation
\reference{} Hunter, D.\ A., \& Gallagher, J.\ S.\ 1985, AJ, 90, 1789
\reference{} Hunter, D.\ A.\ 1997, PASP, 109, 937
\reference{} Hunter, D.\ A., \& Hoffman, L.\ 1999, AJ, 117, 2809
\reference{} Hunter, D.\ A., Wilcots, E.\ M., van Woerden, H., 
Gallagher, J.\ S.,
Kohle, S.\ 1998, ApJ, 495, L47
\reference{} Rao, S., Turnshek, D., \& Briggs, F.\ ApJ, 449, 488
\reference{} Salpeter, E.\ E.\ 1955, ApJ, 121, 161
\reference{} Schneider, S.\ E., Thuan, T.\ X., Magri, C., \& Wadiak, J.\ E.\
1990, ApJS, 72, 245
\reference{} Swaters, R.\ 1999, PhD thesis, Rijksuniversiteit Groningen
\reference{} van den Bergh, S.\ 1988, PASP, 100, 344
\reference{} van Zee, L., Haynes, M.\ P., \& Giovanelli, R.\ 1995, AJ,
109, 990
\reference{} Wilcots, E.\ M., \& Hunter, D.\ A.\ 2001, in preparation
\reference{} Wilcots, E.\ M., \& Miller, B.\ W.\ 1998, ApJ, 116, 2363

\end{references}
\end{document}